# A Thermodynamic Analysis of Enhanced Metastability in Isochoric Supercooled Liquids

**Boris Rubinsky**

*Department of Mechanical Engineering, University of California Berkeley, Berkeley CA 94720*

rubinsky@berkeley.edu

Abstract

Experiments show that isochoric (constant-volume) conditions enhance supercooling stability relative to isobaric (constant-pressure) conditions. Here, combining Helmholtz equilibrium thermodynamics with a first-order perturbation methodology, we derive an inequality governing nucleation stability under volumetric constraint. The derivation provides a general thermodynamic proof that for any substance undergoing phase transformation in which the solid is less dense than the liquid, $v_s > v_l$ ($v_s$ and $v_l$ are the specific volumes of solid and liquid), the Helmholtz driving force for solidification in isochoric systems is smaller than the Gibbs driving force in isobaric systems. Since nucleation rates depend exponentially on the inverse square of the driving force, this provides a thermodynamic basis for the observed suppression of nucleation rates. While a full stochastic treatment is beyond the scope of this work, the reduction in driving force implies a weakening of the bias toward growth of pre-critical fluctuations, increasing their probability of thermal dissolution. The analysis yields a dimensionless isochoric stability number, $\Pi_0 = (v_s - v_l)^2/(L_f \cdot \kappa_l \cdot v_l)$, ($L_f$ - latent heat of fusion, $\kappa_l$ - liquid isothermal compressibility). This number is computable from bulk thermodynamic data alone and provides a geometry-independent criterion for comparing metastable liquid stability across materials and conditions.

**Keywords**: supercooling, isochoric, cryopreservation, Helmholtz, thermodynamic model

## 1. Introduction

Preservation of biological matter at subfreezing temperatures is limited by ice formation, which damages cells and tissue. Supercooling, maintaining water in the liquid phase below its equilibrium melting point, offers a route to subfreezing preservation without freezing damage and is being actively studied for medical, biotechnological, [1,3,22] and food [15,21] applications. Supercooling is, however, a metastable state: the liquid is thermodynamically unstable against ice nucleation, and a variety of strategies have been developed to suppress nucleation, including surface sealing with immiscible liquids [2,11,28], electromagnetic fields [12,13,27], antifreeze proteins [14,29].

An important newly identified strategy is supercooling in an isochoric (constant-volume) confinement. The first comparative study showed that isochorically supercooled aqueous solutions withstand ultrasonication, drop impact, vibration, and thermal cycling more effectively than isobarically supercooled solutions, whether held in open vessels or in surface-sealed isobaric containers [19]. Subsequent work confirmed the enhanced stability for a range of aqueous solutions [7,8] and across different cooling rates in [9,10]. The robustness of isochoric supercooling has also been demonstrated under takeoff conditions[26] in the context of sending supercooled organoids to space [23,24], and the approach has been applied clinically to the preservation of human cardiac microtissues [18] and of an entire pig liver [6,16] and for food preservation[4] .

Despite extensive experimental evidence that isochoric supercooling is more stable than isobaric supercooling, a general thermodynamic explanation for this difference has not been established. The central question addressed in this work is therefore: *why does isochoric confinement intrinsically suppress nucleation relative to isobaric conditions?* The common framework for nucleation kinetics is classical nucleation theory (CNT), in which the nucleation rate $J \propto \exp(-\Delta G^*/k_B T)$ with $\Delta G^* \propto \gamma^3/(\Delta g)^2$, where $\Delta G^*$ is the free-energy barrier to nucleation, $k_B$ is the Boltzmann constant, T is temperature, $\gamma$ is the solid–liquid interfacial energy, and $\Delta g$ is the bulk free-energy difference per unit volume between liquid and solid. The nucleation barrier therefore scales as the inverse square of $\Delta g$. The present paper introduces an analytical approach to explaining the increased stability of isochoric supercooled solutions in which the solid is less dense than the liquid upon phase transformation, relative to supercooling in an isobaric system. Rather than tracking nucleation kinetics as a function of absolute system volume or nucleus size, we analyze the thermodynamic driving forces directly, comparing the Helmholtz driving force $|\Delta f_{isoV}|$ with the Gibbs driving force $|\Delta g_{isoP}|$ using a first-order perturbation in the nascent solid fraction $\xi$.

Here we combine two analytical frameworks to derive a thermodynamic inequality governing nucleation stability under constant volume constraint and constant pressure constraint, for any substance in which the solid is less dense than the liquid ($v_s > v_l$ where $v_s$ and $v_l$ are the specific volumes of solid and liquid). Their combination shows a self-limiting thermodynamic feedback absent under isobaric conditions: for any substance in which the solid is less dense than the liquid, the Helmholtz driving force for solidification in an isochoric system is strictly less than the Gibbs driving force in an isobaric system. This approach yields three general results: (i) a thermodynamic proof that $|\Delta f_{isoV}| < |\Delta g_{isoP}|$ for any substance with a negative Clapeyron slope, regardless of system geometry or absolute volume; (ii) the explicit mechanism linking this inequality to Helmholtz convexity and the Clapeyron relation through a self-limiting thermodynamic feedback; and (iii) a dimensionless isochoric stability number $\Pi_0 = (v_s - v_l)^2 / (L_f \cdot \kappa_l \cdot v_l)$ (where $L_f$ is the latent heat of fusion and $\kappa_l$ is the thermal compressibility of the liquid), that characterizes the thermodynamic tendency for nucleation suppression as an intrinsic material property, independent of system geometry. The central result of this work is that isochoric supercooling is intrinsically more stable because the thermodynamic driving force for solidification is reduced by a self-limiting pressure–temperature feedback that does not exist under isobaric conditions. Beyond its implications for the classical nucleation barrier, this result also provides insight into the stochastic pre-critical nucleation process, showing that isochoric conditions reduce the thermodynamic bias that drives fluctuations toward growth.

While the analysis was motivated by isochoric supercooling of aqueous solutions, the derivation is general: it applies to any substance whose solid is less dense than its liquid under the phase transition conditions considered, including silicon, gallium, bismuth, germanium, and antimony - materials of broad interest in semiconductor processing, precision casting, and materials science.

## 2. Thermodynamic Analysis

Throughout this paper, g and f denote specific Gibbs and Helmholtz free energies (per unit mass); subscripts $l$ and $s$ denote liquid and solid phases; subscripts $isoP$ and $isoV$ denote isobaric and isochoric conditions; $L_f$ is the latent heat of fusion per unit mass; $T_m(P_0)$ is the equilibrium melting temperature at the reference pressure $P_0$; $\xi \ll 1$ is the nascent solid mass fraction used as perturbation parameter; $v_s$ and $v_l$ are the specific volumes of the solid and liquid phases; $\kappa_l$ is the isothermal compressibility of the liquid phase at reference pressure $P_0$. Water and ice Ih are used as a concrete illustration throughout; all results apply without modification to any substance with $v_s > v_l$ (negative Clapeyron slope), including silicon, gallium, bismuth, germanium, and antimony.

Following is a brief description of the isochoric freezing process, to aid in the understanding of the analysis. Freezing of aqueous biological systems is conventionally carried out under isobaric conditions, where the phase transition occurs near atmospheric pressure and the equilibrium melting temperature is determined primarily by solute concentration. Isochoric (constant-volume) freezing produces fundamentally different thermodynamic states in which pressure and phase composition are coupled through the fixed-volume constraint [20,25]. When a water-filled rigid chamber is cooled below the equilibrium melting temperature, formation of ice Ih, which has a lower density than water, generates elevated hydrostatic pressure that resists the volume expansion associated with solidification. As the system temperature is lowered, the system evolves along the water–ice Ih liquidus, with both pressure and phase composition determined by the fixed-volume condition (Figure 1) [5,25].

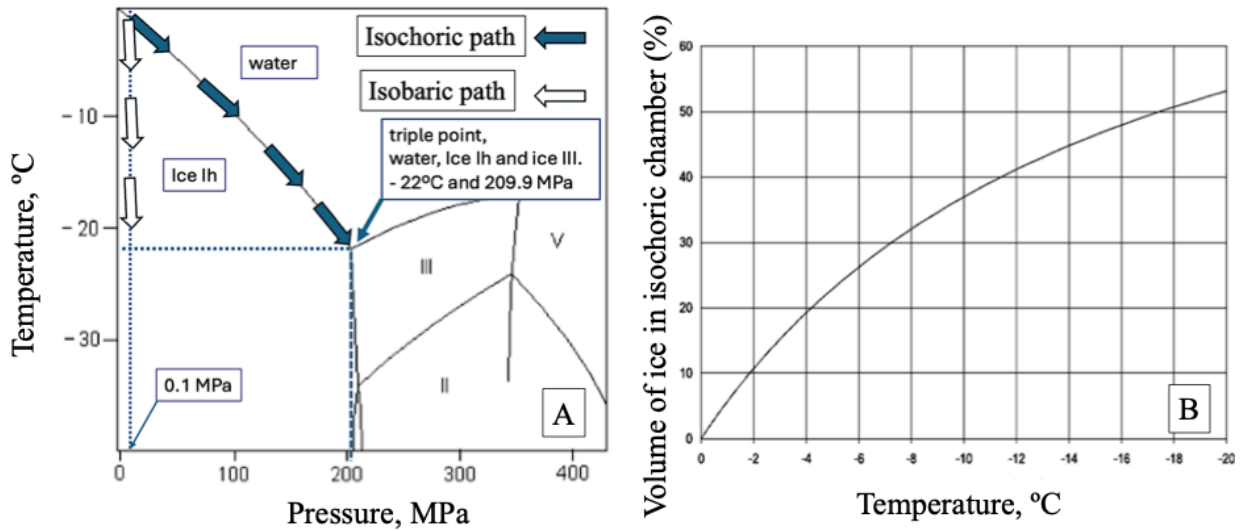

*Figure 1.* Thermodynamic basis of isochoric freezing. (A) Pressure–temperature phase diagram showing the isochoric process path along the water–ice $I_h$ liquidus to −21.9°C, 209.9 MPa at the triple point. The negative Clapeyron slope

reflects the fact that ice has a larger specific volume than liquid water. (B) Ice volume and mass fraction as functions of temperature depression; complete solidification is thermodynamically prohibited, a liquid fraction always persists. (Reprinted from: B. Rubinsky, P.A. Perez, and E.M. Carlson, "The thermodynamic principles of isochoric cryopreservation," Cryobiology 50(2), 121–138 (2005). Figure 7. With Permission from Elsevier)

The natural thermodynamic potential for this system is the Helmholtz free energy F(T,V)[20], in which pressure emerges as a derived quantity from the common tangent to the free-energy surfaces was developed first by Powell-Palm and colleagues[17,20] (Figure 2A). The two-phase coexistence region appears as a finite area in T–V space, with phase fractions given by the Lever rule (Figure 1C) [17]. This T–V framework is the natural one for the present problem, because nucleation corresponds to a horizontal displacement across pressure isoclines, a geometric operation whose spatial simultaneity is not apparent in the conventional T–P representation.

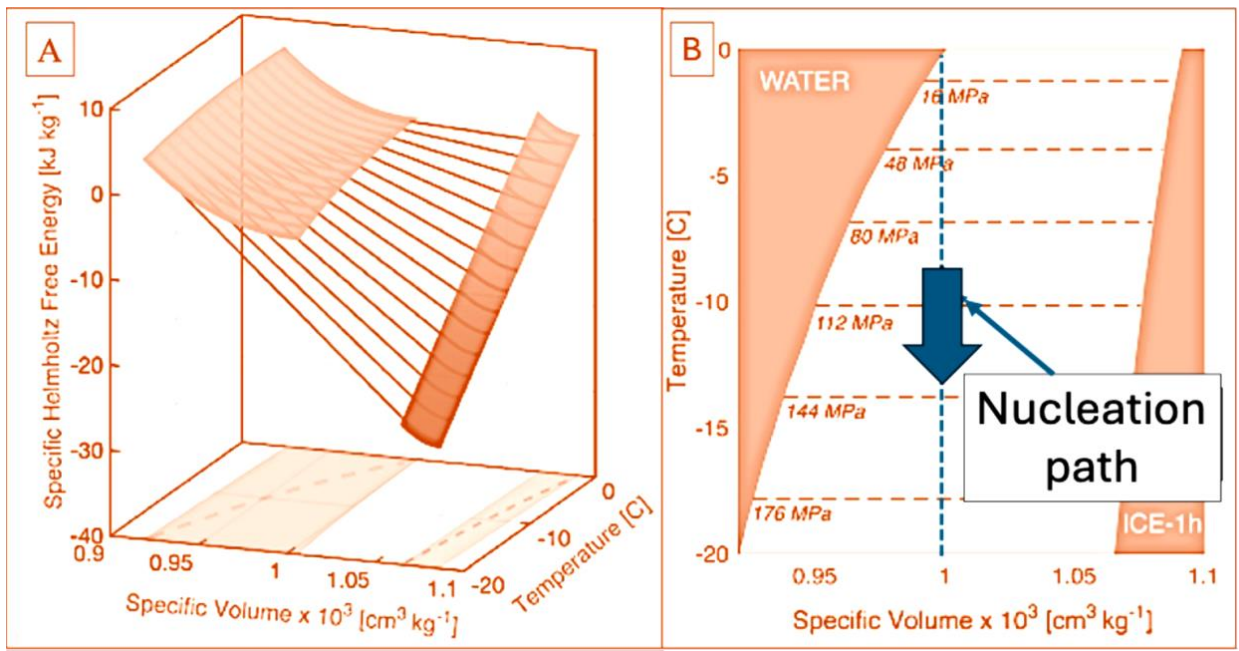

*Figure 2.* Helmholtz free energy *T–V* phase diagram of water. (A) Free energy surfaces of liquid water and ice I$_h$ as functions of temperature and specific volume; the common tangent construction (Lines of Dissipated Energy) gives the equilibrium pressure. (B) Projected *T–V* phase diagram with broad two-phase equilibrium region traversed by pressure isoclines. Nucleation in an isochoric system traces a vertical path; the equilibrium phase fraction is determined by the Lever rule. The arrow shows a slow equilibrium decompression path. (Modified from Powell-Palm, M.J., Rubinsky, B. & Sun, W.[20] Freezing water at constant volume and under confinement. *Commun Phys* **3**, 39 (2020). https://doi.org/10.1038/s42005-020-0303-9 under the Terms of Creative Commons CC BY)

https://rdcu.be/faEjW)

## 2a. Free-Energy Geometry and the Isobaric Baseline

The thermodynamic driving force for solidification nucleation is the difference in specific free energy between the liquid and solid phases evaluated at the conditions prevailing at the moment of nucleation. The choice of thermodynamic potential is determined by the constraints imposed on the system.

For a system held at constant pressure $P$, the natural potential is the Gibbs free energy $G(T, P)$, whose surfaces are concave downward in both $T$ and $P$. For a system held at constant volume $V$, the natural potential is the Helmholtz free energy $F(T, V)$, whose surfaces are concave downward in $T$ but convex in $V$. This convexity in volume is the geometric origin of the isochoric stabilisation effect.

Under constant-pressure conditions, phase coexistence is governed by the intersection of the Gibbs surfaces of the liquid and solid phases. Along the melting line:

$$\Delta g(T_m(P), P) \equiv g_l(T_m(P), P) - g_s(T_m(P), P) = 0 \quad (1)$$

The Gibbs differential at constant pressure is $d(\Delta g) = -\Delta s\, dT$, with $\Delta s = L_f/T_m$. A first-order expansion about the reference melting point $T_m(P_0)$ gives:

$$\Delta g_{isoP} = (L_f/T_m)(T_m(P_0) - T) = (L_f/T_m)\Delta T_{isoP} \quad (2)$$

where $\Delta T_{isoP} = T_m(P_0) - T$ is the imposed supercooling. The magnitude of the isobaric driving force grows linearly with supercooling. Geometrically, $\Delta g$ measures the vertical separation between the two Gibbs surfaces at $(T, P)$; below the melting point this gap opens linearly with supercooling.

The Helmholtz free energy is related to the Gibbs free energy by the Legendre transform:

$$F(T, V) = G(T, P(V)) - PV \quad (3)$$

Because $V$ is an extensive variable, the Helmholtz surface $F(V)$ is convex: $\partial^2 F/\partial V^2 > 0$. The crucial thermodynamic identity is:

$$(\partial F/\partial V)_T = -P \quad (4)$$

Under isobaric conditions, the pressure is externally fixed and the slope of $F(V)$ is clamped by the pressure reservoir. The convexity of $F(V)$ is present but never engaged: $\Delta P = 0$ identically, the liquidus is never shifted, and the full Gibbs driving force $\Delta g_{isoP}$ remains available throughout the process. The isochoric case is the opposite limit, derived in the following sections.

## 2b. Isochoric Pressure Rise from Helmholtz Convexity

In an isochoric system, the natural potential is $F(T, V)$ with the slope condition (4). Consider a rigid container filled with liquid water at temperature $T < T_m(P_0)$, held in a metastable supercooled state. Before any ice forms, the system occupies a single point on the $F_l(V)$ curve at the container's specific volume $V_0$, and the prevailing pressure is:

$$P_0 = -(\partial F_l/\partial V)_{T_0} \quad (5)$$

Now let an infinitesimal ice fraction $\xi \ll 1$ form. Because ice Ih has a larger specific volume than liquid water ($v_s > v_l$), formation of any ice fraction tends to expand the mixture. In a rigid container this expansion is forbidden; the container exerts a restoring pressure on both phases, compressing the mixture back to $V_0$.

In the Helmholtz picture this constraint is captured by the parallel tangent construction introduced by Powell-Palm et al[20]: the specific volumes and free energies of the two coexisting phases are identified by points on $F_l(V)$ and $F_s(V)$ whose tangent lines are parallel (equal pressures in the two phases, as required by mechanical equilibrium). As the ice fraction $\xi$ grows from zero, the common slope of these parallel tangents changes continuously, tracking the evolving system pressure.

At solid fraction $\xi$, the mixture specific volume is fixed at $V_0$ by the lever rule:

$$V_0 = (1 - \xi)v_l(\xi) + \xi v_s(\xi) \quad (6)$$

where $v_l(\xi)$ and $v_s(\xi)$ are the specific volumes of the liquid and solid phases, determined jointly by the parallel tangent condition and volume conservation. The parallel tangent condition requires equal pressure in both phases:

$$-(\partial F_l/\partial v)v_l = -(\partial F_s/\partial v)v_s = P(\xi) \quad (7)$$

To make the pressure-rise derivation explicit, we differentiate the lever rule (6) with respect to $\xi$ and evaluate at $\xi = 0$. With the compressibility relation $dv/dP = -\kappa_l v_l$ and the equal-pressure condition (7), this gives:

$$0 = (v_{s0} - v_{l0}) + dv_l/d\xi_{\xi=0} \quad (8)$$

where $v_{s,0}$ and $v_{l,0}$ are the pure-phase specific volumes at $P_0$. Equation (8) confirms that the liquid specific volume must decrease as $\xi$ increases, exactly compensating the volume gained by forming the less-dense solid—which is equivalent to a pressure rise, as derived below.

The pressure rise is obtained via a direct volume-conservation argument. If pressure did not change, phase conversion would increase the mixture specific volume by:

$$\Delta v_{phase} = \xi(v_s - v_l) \quad (9)$$

To maintain constant total volume, this expansion must be exactly offset by elastic compression of the liquid. To first order in $\xi$ (so that the compressible fraction is effectively the whole liquid volume), the compression is:

$$\Delta v_{comp} = -\kappa_l v_l \Delta P \quad (10)$$

The approximation of compressing the full liquid volume $v_l$ rather than the remaining fraction $(1-\xi)v_l$ introduces an error of order $O(\xi^2)$, which is negligible at the order of the present calculation.

Volume conservation requires $\Delta v_{phase} + \Delta v_{comp} = 0$. Substituting (9) and (10) and solving for $\Delta P$:

$$\Delta P = \xi(v_s - v_l)/(\kappa_l v_l) + O(\xi^2) \quad (11)$$

Since $(v_s - v_l) > 0$ and $\kappa_l > 0$, we have $\Delta P > 0$ for any $\xi > 0$: the system is self-pressurizing. This is a direct consequence of the convexity of $F(V)$: forming a small solid fraction forces the slope $\partial F/\partial V$ to become more negative (steeper parallel tangent), which by Eq. (4) means the pressure $P = -\partial F/\partial V$ increases.

## 2c. Melting-Line Depression via the Clapeyron Relation

The equilibrium melting temperature is read from the T–V phase diagram by projecting the lowest-energy convex hull of the two $F(V)$ curves onto the T–V plane. The boundary of the two-phase coexistence region at any temperature is set by the common tangent between the two F(V) curves; the slope of that tangent $\partial F/\partial V$, equals $-P_{eq}(T)$, the negative of the equilibrium melting pressure at temperature T. The pressure perturbation $\Delta P = O(\xi)$ from Eq. (11) shifts the system along this melting line.

Expanding $T_m$ about $P_0$ to first order in $\Delta P$:

$$T_m(P_0 + \Delta P) = T_m(P_0) + (dT_m/dP)_{P_0}\Delta P + O(\Delta P^2) \quad (12)$$

Since $\Delta P = O(\xi)$, the term $O(\Delta P^2) = O(\xi^2)$ and is dropped at first order. The Clapeyron relation gives the slope of the melting curve:

$$dT_m/dP = T_m(v_l - v_s)/L_f \quad (13)$$

For ice Ih, $v_s > v_l$, so $dT_m/dP < 0$. This negative slope is the anomalous property shared by water, silicon, gallium, bismuth, and related materials. In the $T - V$ phase diagram, this negativity is directly visible as the water-side boundary of the two-phase coexistence region sloping to the left (lower temperatures) as temperature decreases, reflecting that the common tangent becomes steeper (more negative slope, higher pressure) at lower temperatures. The pressure rise $\Delta P$ therefore moves the system to the left along the liquidus, lowering the effective melting temperature.

## 2d. Effective Supercooling Under Isochoric Constraint

The effective thermodynamic supercooling under isochoric conditions is the gap between the shifted melting temperature and the imposed system temperature:

$$\Delta T_{isoV} = T_m(P_0 + \Delta P) - T \quad (14)$$

Here $T$ is the externally imposed temperature of the rigid container—a fixed boundary condition, not a perturbation variable. The perturbative expansion is carried out solely in the small ice fraction $\xi$.

Substituting (12) into (14) and noting that $T_m(P_0) - T = \Delta T_{isoP}$ (the isobaric supercooling at the same temperature, T):

$$\Delta T_{isoV} = \Delta T_{isoP} + (dT_m/dP)_{P_0}\Delta P + O(\xi^2) \quad (15)$$

Inserting Eq. (11) for $\Delta P$ and Eq. (13) for $dT_m/dP$:

$$\Delta T_{isoV} = \Delta T_{isoP} + [T_m(v_l - v_s)/L_f] \cdot [\xi(v_s - v_l)/(\kappa_l v_l)] + O(\xi^2) \quad (16)$$

Evaluating the product of the two bracketed terms:

$$[T_m(v_l - v_s)/L_f] \cdot [\xi(v_s - v_l)/(\kappa_l v_l)] = -T_m \xi (v_s - v_l)^2/(L_f \kappa_l v_l) \quad (17)$$

This term is strictly negative since $T_m$, $\xi$, $(v_s - v_l)^2$, $L_f$, $\kappa_l$, and $v_l$ are all positive. Therefore:

$$\Delta T_{isoV} = \Delta T_{isoP} - T_m \xi (v_s - v_l)^2/(L_f \kappa_l v_l) + O(\xi^2) \quad (18)$$

Since the second term on the right is strictly positive (subtracted), to first order in $\xi$:

$$\Delta T_{isoV} < \Delta T_{isoP} \quad (19)$$

The Helmholtz-slope argument makes the mechanism transparent: the pressure rise ΔP steepens the common tangent (makes its slope more negative), which by the geometry of the T–V phase diagram corresponds to moving down and to the left along the liquidus, reducing $T_m$. Since T is fixed, this reduction closes the effective supercooling gap. Under isobaric conditions none of this feedback operates: the liquidus is never shifted, and the full imposed supercooling $\Delta T_{isoP}$ remains available as driving force throughout.

## 2e. Isochoric Driving Force and the Central Inequality

The bulk free-energy driving force for solidification under isochoric conditions is the Helmholtz free-energy difference between liquid and solid evaluated at the effective supercooling $\Delta T_{isoV}$. The expansion is identical in structure to the isobaric case in Section 3, but the Helmholtz potential is used and the expansion is about the pressure-shifted melting point:

$$\Delta f_{isoV} = (L_f/T_m)\Delta T_{isoV} + O(\xi^2) \quad (20)$$

The coefficient $L_f/T_m$ in Eq. (20) is the same as in the isobaric case Eq. (2). This follows from the fundamental Helmholtz relation $dF = -S\,dT - P\,dV$ which gives $(\partial f/\partial T)_V = -s$ at constant volume. The temperature derivative of the driving force $\Delta f = f_l - f_s$ at constant volume is therefore $(\partial \Delta f/\partial T)_V = -(s_l - s_s) = -\Delta s = -L_f/T_m$, which is the same coefficient that appears in the isobaric case The only difference between Eq. (20) and Eq. (2) is the expansion point: Eq. (2) expands about $T_m(P_0)$, while Eq. (20) expands about the pressure-shifted melting point $T_m(P_0 + \Delta P)$ derived in Section 5. No separate Legendre correction is needed because the shift of the expansion point already accounts for the pressure change.

$$|\Delta f_{isoV}| < |\Delta g_{isoP}| + O(\xi^2) \quad (21)$$

Since $L_f/T_m > 0$, substitution of inequality (19) into Eq. (20) and comparison with Eq. (2) yields Eq. (21). To first order in the solid fraction, the magnitude of the bulk free-energy driving force under constant-volume conditions is strictly smaller than that under constant-pressure conditions for any system with a negative Clapeyron slope.

This result establishes a general thermodynamic ordering between the Helmholtz and Gibbs driving forces. Its implications for nucleation kinetics and metastable behavior are examined in Sections 3 and 4.

The Helmholtz geometry makes the three-step self-limiting feedback self-contained and directly readable from the T–V phase diagram:

**(i) Convexity of F(V)** generates $\Delta P = O(\xi) > 0$ [Eq. (11)]: any solid formation forces $\partial F/\partial V$ more negative, identically a pressure rise.

**(ii) Negative Clapeyron slope** converts $\Delta P$ into a reduction of $T_m$ [Eqs. (12)–(13)]: the steeper common tangent in T–V space moves the equilibrium melting point downward and to the left.

**(iii) Fixed imposed temperature T** means the gap $\Delta T_{isoV} = T_m(P_0 + \Delta P) - T$ is smaller than $\Delta T_{isoP} = T_m(P_0) - T$, reducing the free-energy driving force.

The isobaric case is the degenerate limit in which the slope of F(V) is clamped externally: $\Delta P = 0$ identically, the liquidus is never shifted, and the full Gibbs driving force is recovered. The isochoric result is not a perturbative correction to the isobaric case but a consequence of allowing the thermodynamic potential geometry to respond to phase change.

## 2f. The Dimensionless Isochoric Stability Number $\Pi_0$

Equation (18) shows that the reduction in effective supercooling takes the form:

$$\Delta T_{isoV} = \Delta T_{isoP} - \Pi_0 \cdot T_m \cdot \xi + O(\xi^2) \quad (22)$$

where the dimensionless isochoric stability number is:

$$\Pi_0 = (v_s - v_l)^2 / (L_f \cdot \kappa_l \cdot v_l) \quad (23)$$

Dimensional consistency: $(v_s - v_l)^2$ has units (m³ kg⁻¹)²; $[L_f]$ has units J kg⁻¹ = m² s⁻²; $[\kappa_l]$ has units Pa⁻¹ = m s² kg⁻¹; $[v_l]$ has units m³ kg⁻¹. The denominator product has units m⁶ kg⁻², identical to the numerator, confirming $\Pi_0$ is dimensionless.

Dividing Eq. (22) by $T_m$ yields the most compact dimensionless form:

$$(\Delta T_{isoP} - \Delta T_{isoV}) / T_m = \Pi_0 \cdot \xi + O(\xi^2) \quad (24)$$

## 3. Results and Discussion

Experimental studies have consistently shown that isochoric enclosure enhances supercooling stability relative to supercooling in isobaric systems. The present analysis explains why isochoric supercooling is intrinsically more stable than isobaric supercooling. By combining the Helmholtz free-energy framework for constant-volume systems with a first-order perturbation in the nascent solid fraction $\xi$, we derive the inequality $|\Delta f_{isoV}| < |\Delta g_{isoP}|$ - valid for any substance with a negative Clapeyron slope, regardless of system geometry or absolute volume - and extract from it the dimensionless isochoric stability number $\Pi_0$ as thermodynamic quantity.

The derivation reveals a three-step self-limiting feedback that is entirely absent under isobaric conditions. First, the convexity of the Helmholtz free energy F(V) converts any nascent solidification into a pressure rise $\Delta P = \xi(v_s - v_l)/(\kappa_l v_l)$ [Eq. (11)]: because ice is less dense than water, forming a small solid fraction in a rigid container forces the slope $\partial F/\partial V$ to become more negative, which by the identity $(\partial F/\partial V)_T = -P$ is identically a pressure increase. Second, the negative Clapeyron slope $dT_m/dP = T_m(v_l - v_s)/L_f$ converts this pressure rise into a depression of the effective melting temperature [Eqs. (12)–(13)], lowering $T_m$ below its reference value at P₀. Third, because the container temperature T is a fixed external boundary condition, the effective supercooling gap $\Delta T_{isoV} = T_m(P_0 + \Delta P) - T$ strictly smaller than the isobaric supercooling $\Delta T_{isoP} = T_m(P_0) - T$. Under isobaric conditions none of this feedback operates: the slope of F(V) is externally clamped, ΔP = 0 identically, the liquidus is never shifted, and the full Gibbs driving force $\Delta g_{isoP}$ remains available throughout.

Three features of this result deserve emphasis. First, the inequality is geometry-independent: it makes no assumption about confinement geometry, surface chemistry, or absolute system volume, and follows from bulk thermodynamic quantities alone. This is important because the experimental literature cited in the introduction spans a wide range of container geometries, biological matrices, and food systems, yet the enhancement of supercooling stability is observed consistently — precisely the behaviour one expects from a mechanism rooted in bulk thermodynamic driving forces rather than surface or geometric effects. Second, the perturbation variable here is the nascent solid fraction ξ, which is a macroscopic, system-level quantity, rather than the nucleus radius r used in classical nucleation theory; the two perturbation approaches are complementary, and their mutual consistency is addressed in the context of prior work below. Third, because the derivation requires only $v_s > v_l$ and $dT_m/dP < 0$, the result is not restricted to water but applies without modification to any substance with a negative Clapeyron slope.

It is equally important to note what the isochoric constraint does not do: it does not eliminate the thermodynamic driving force for freezing. The imposed supercooling T_m(P₀) − T remains the primary control parameter of thermodynamic instability. The isochoric correction is, to first order in ξ, independent of the depth of supercooling and acts as a uniform offset, the same fractional reduction in effective supercooling applies whether the system is 1 °C or 10 °C below the melting point. This is the physical content of Eq. (24): the dimensionless supercooling reduction ($\Delta T_{isoP} - \Delta T_{isoV}$) / $T_m$ is proportional to ξ with a proportionality constant $\Pi_0$ that depends only on material properties, not on the imposed temperature or system geometry.

Classical nucleation theory expresses the nucleation barrier as the result of competition between interfacial energy and a bulk free-energy driving force. Under isothermal–isobaric conditions, this driving force is expressed in terms of the Gibbs free-energy difference, yielding a barrier of the form $\Delta G^*_{isoP} \propto \gamma^3/(\Delta g_{isoP})^2$. Under isochoric conditions, however, the appropriate thermodynamic potential is the Helmholtz free energy. The structure of the nucleation barrier remains unchanged, but the bulk driving force is replaced by the Helmholtz free-energy difference, giving $\Delta F^*_{isoV} \propto \gamma^3/(\Delta f_{isoV})^2$. Since the present analysis shows that $|\Delta f_{isoV}| < |\Delta g_{isoP}|$, the nucleation barrier under isochoric conditions is larger, providing a thermodynamic explanation for the enhanced stability of isochoric supercooling. Even the modest reduction in $|\Delta f_{isoV}|$ established by Eq. (21) therefore produces a disproportionately large suppression of J, amplifying a linear reduction in driving force into an exponential suppression of nucleation rate. While the bulk thermodynamic driving force that enters classical nucleation theory is modified, the interfacial energy and nucleus geometry is unchanged. The present formulation is therefore complementary to classical nucleation theory and provides a thermodynamic basis for the observed reduction in nucleation rates under isochoric conditions.

In addition to increasing the classical nucleation barrier, the inequality provides a thermodynamic description of pre-critical nucleation dynamics. It has implications beyond the classical nucleation barrier and extends to the stochastic dynamics of pre-critical fluctuations. A nascent solid embryo does not evolve deterministically but undergoes thermally driven fluctuations in size, with growth favored by the bulk thermodynamic driving force and dissolution favored by thermal agitation. The net tendency for growth is therefore governed by the ratio between the bulk driving force and thermal fluctuations.

Because the present analysis shows that $|\Delta f_{\text{isoV}}| < |\Delta g_{\text{isoP}}|$, the thermodynamic bias favoring growth is reduced under isochoric conditions at all embryo sizes. As a consequence, even before the critical nucleus size is reached, fluctuations are more likely to dissolve and less likely to grow. Equivalently, the free-energy landscape governing embryo evolution is shallower under isochoric constraint, reducing the deterministic drift toward growth while leaving the magnitude of thermal fluctuations unchanged. This effect is not explicitly captured within classical nucleation theory, which focuses on the height of the critical barrier. The results of this analysis reflect a more general weakening of the thermodynamic bias toward solidification. Isochoric confinement therefore suppresses nucleation not only by increasing the nucleation barrier, but also by reducing the probability that pre-critical fluctuations will reach that barrier. This provides a deeper thermodynamic explanation for the experimentally observed stability of isochoric supercooling: the system is stabilized both by an increased barrier and by a reduced tendency of fluctuations to grow toward that barrier.

The analysis provides a thermodynamic, geometry-independent explanation for the experimental observations described in the introduction: that isochorically supercooled systems resist nucleation induced by mechanical and biological stimuli across diverse container sizes, biological matrices, and food systems. The result does not invoke nucleation site models, surface effects, or knowledge of system geometry, and is therefore consistent with the broad and varied experimental record.

The dimensionless isochoric stability number $\Pi_0 = (v_s - v_l)^2 / (L_f \cdot \kappa_l \cdot v_l)$ encodes the strength of this feedback as a single material property. Its structure is physically transparent. The density contrast $(v_s - v_l)$ appears squared because it enters the feedback twice: once in generating the pressure rise $\Delta P$ via volume conservation [Eq. (11)], and once in the Clapeyron slope $dT_m/dP$ [Eq. (13)], which also depends on $(v_s - v_l)$. A larger latent heat $L_f$ yields a shallower Clapeyron slope and weaker melting-point depression, so $1/L_f$ appears in the denominator. A less compressible liquid $\kappa_l$ generates a larger pressure rise per unit solid fraction, strengthening the feedback, so $1/\kappa_l$ also appears. All quantities are measurable from bulk thermodynamic data at the melting point, with no geometric parameters.

The inequality Eq. (21) and the stability number $\Pi_0$ hold for any substance with $dT_m/dP < 0$. This class includes, beyond water and cryopreservation solutions, silicon (semiconductor solidification under rigid mould confinement), bismuth and gallium (precision casting), germanium and antimony, and water in rigid porous matrices (atmospheric aerosol nucleation, geological frost-heave). In principle, $\Pi_0$ provides a framework for comparing materials with respect to isochoric nucleation suppression based on thermodynamic properties alone. The combination of Helmholtz geometry with first-order perturbation theory used here may also serve as a broader methodological template for studying metastable liquid stability under volumetric confinement; natural extensions include multicomponent solutions relevant to biological cryopreservation, finite solid fractions beyond the first-order approximation, and non-uniform pressure distributions arising from compliant vessel walls.

# Conclusion

We have shown that isochoric supercooling is intrinsically more stable than isobaric supercooling by deriving a general thermodynamic inequality using the Helmholtz free-energy framework and a first-order perturbation in the nascent solid fraction. The result holds for any substance with a negative Clapeyron slope and is independent of system geometry.

The derivation reveals a three-step self-limiting feedback: Helmholtz convexity generates a pressure rise, the negative Clapeyron slope converts this into a depression of the effective melting temperature, and the fixed container temperature closes the effective supercooling gap — that is absent under isobaric conditions. Through the inverse-square dependence of the CNT nucleation barrier on driving force, even the modest reductions quantified here produce disproportionately large suppression of nucleation rates, consistent with experiment.

In addition, the analysis provides a thermodynamic interpretation of pre-critical nucleation dynamics, demonstrating that isochoric conditions reduce the bias that drives fluctuations toward growth, thereby decreasing the probability that such fluctuations will evolve into critical nuclei.

The analysis yields a dimensionless isochoric stability number $\Pi_0 = (v_s - v_l)^2 / (L_f \cdot \kappa_l \cdot v_l)$, computable from bulk thermodynamic data alone, that provides a geometry-independent framework for comparing isochoric nucleation suppression across materials and conditions. This provides a rigorous, geometry-independent thermodynamic basis for the experimentally documented isochoric supercooling enhancement, and a practical dimensionless design criterion for applications ranging from cryopreservation to semiconductor solidification.

## Data Availability

The analytical derivations are contained within the paper. No experimental data were generated in this study.

## Competing Interests

B.R. holds a minority share in BioChoric Inc., which commercializes isochoric technology.

## Funding

NSF ERC ATP-BIO Grant No. 1941543.

## Author Contributions

B.R. conceived the study, developed the analysis, and wrote the paper.

## Acknowledgements

The author used Claude (Anthropic) to assist with editing, linguistic refinement, computation of $\Pi_0$, and document preparation. The final version was reviewed and approved by the author.